\documentclass[aps,pre,twocolumn,superscriptaddress,floatfix]{revtex4}

\usepackage{graphicx,bm,amssymb,amsmath,dcolumn,hyperref}
\usepackage{multirow}
\usepackage{enumerate}
\usepackage{enumitem}
\usepackage{color}
\usepackage{subfigure}  %added.
\usepackage{epstopdf}
\usepackage{bbm}
\usepackage{graphicx}
\usepackage{hyperref}
\usepackage{threeparttable}
\usepackage{amsthm}
\usepackage{mathtools}
\usepackage{color}
\usepackage{tikz}
\usepackage{float}

\begin{document}
\newcommand{\be}{\begin{equation}}
\newcommand{\ee}{\end{equation}}
\newcommand{\half}{\frac{1}{2}}
\newcommand{\ith}{^{(i)}}
\newcommand{\im}{^{(i-1)}}
\newcommand{\gae}
{\,\hbox{\lower0.5ex\hbox{$\sim$}\llap{\raise0.5ex\hbox{$>$}}}\,}
\newcommand{\lae}
{\,\hbox{\lower0.5ex\hbox{$\sim$}\llap{\raise0.5ex\hbox{$<$}}}\,}

\definecolor{blue}{rgb}{0,0,1}
\definecolor{red}{rgb}{1,0,0}
\definecolor{green}{rgb}{0,1,0}
\newcommand{\blue}[1]{\textcolor{blue}{#1}}
\newcommand{\red}[1]{\textcolor{red}{#1}}
\newcommand{\green}[1]{\textcolor{green}{#1}}
\newcommand{\orange}[1]{\textcolor{orange}{#1}}
\newcommand{\yd}[1]{\textcolor{blue}{#1}}

\newcommand{\scrA}{{\mathcal A}}
\newcommand{\scrE}{{\mathcal E}} %added.
\newcommand{\scrF}{{\mathcal F}} %added.
\newcommand{\scrL}{{\mathcal L}}
\newcommand{\scrM}{{\mathcal M}} %added.
\newcommand{\scrN}{{\mathcal N}}
\newcommand{\scrS}{{\mathcal S}}
\newcommand{\scrs}{{\mathcal s}}
\newcommand{\scrP}{{\mathcal P}}
\newcommand{\scrO}{{\mathcal O}}
\newcommand{\scrR}{{\mathcal R}}
\newcommand{\scrC}{{\mathcal C}}
\newcommand{\scrV}{{\mathcal V}}
\newcommand{\scrD}{{\mathcal D}}
\newcommand{\PP}{\mathbb{P}}
\newcommand{\ZZ}{\mathbb{Z}}
\newcommand{\EE}{\mathbb{E}}
\renewcommand{\d}{\mathrm{d}}
\newcommand{\dm}{d_{\rm min}}
\newcommand{\rhojunction}{\rho_{\rm j}}
\newcommand{\rhojunctionLim}{\rho_{{\rm j},0}}
\newcommand{\rhobranch}{\rho_{\rm b}}
\newcommand{\rhobranchLim}{\rho_{{\rm b},0}}
\newcommand{\rhononbridge}{\rho_{\rm n}}
\newcommand{\rhononbridgeLim}{\rho_{{\rm n},0}}
\newcommand{\percolationCluster}{C}
\newcommand{\leafFreeCluster}{C_{\rm \ell f}}
\newcommand{\bridgeFreeCluster}{C_{\rm bf}}
\newcommand{\df}{d_{\rm f}}
\newcommand{\yt}{y_{\rm t}}
\newcommand{\bfk}{{\bf k}}
\newcommand{\ytstar}{y^*_{\rm t}}
\newcommand{\yh}{y_{\rm h}}
\newcommand{\yhstar}{y^*_{\rm h}}
\newcommand{\dfprime}{d'_{\rm f}}
\newcommand{\tildechiprimeL}{\tilde{\chi}^\prime_{\rm l}}
\newcommand{\tildechiprimeH}{\tilde{\chi}^\prime_{\rm h}}
\newcommand{\gammaH}{\gamma^\prime_{\rm h}}
\newcommand{\gammaL}{\gamma^\prime_{\rm l}}

\makeatletter
\newcommand{\rmnum}[1]{\romannumeral #1}
\makeatother

%added.
%\newcommand{\dm}{d_{\rm min}}

\title{Complete graph and Gaussian fixed point asymptotics in the five-dimensional Fortuin-Kasteleyn Ising model with periodic boundaries}
\date{\today}
\author{Sheng Fang}
\affiliation{Hefei National Laboratory for Physical Sciences at Microscale and 
Department of Modern Physics, University of Science and Technology of China, 
Hefei, Anhui 230026, China}
\author{Jens Grimm}
\affiliation{ARC Centre of Excellence for Mathematical and Statistical Frontiers (ACEMS),
School of Mathematics, Monash University, Clayton, Victoria 3800, Australia}
\author{Zongzheng Zhou}
\email{eric.zhou@monash.edu}
\affiliation{ARC Centre of Excellence for Mathematical and Statistical Frontiers (ACEMS),
School of Mathematics, Monash University, Clayton, Victoria 3800, Australia}
\author{Youjin Deng}
\email{yjdeng@ustc.edu.cn}
\affiliation{Hefei National Laboratory for Physical Sciences at Microscale and 
Department of Modern Physics, University of Science and Technology of China, 
Hefei, Anhui 230026, China}
\affiliation{CAS Center for Excellence and Synergetic Innovation Center in Quantum Information and Quantum Physics, University of Science and Technology of China, Hefei, Anhui 230026, China}

\begin{abstract}
We present an extensive Markov-chain Monte Carlo study of the finite-size scaling behavior of the Fortuin-Kasteleyn Ising model on five-dimensional hypercubic lattices with periodic boundary conditions. We observe that physical quantities, which include the contribution of the largest cluster, exhibit complete graph asymptotics. However, for quantities, where the contribution of the largest cluster is removed, we observe that the scaling behavior is mainly controlled by the Gaussian fixed point. Our results therefore suggest that \textit{both} scaling predictions, i.e.~the complete graph \textit{and} the Gaussian fixed point asymptotics, are needed to provide a complete description for the five-dimensional finite-size scaling behavior on the torus.\end{abstract}
%\pacs{05.50.+q, 05.70.Jk, 64.60.F-}
\maketitle

\section{Introduction}
\label{Introduction}

Models of critical phenomena typically possess an upper critical dimension $d_{\rm c}$, such that in dimensions $d\ge d_{\rm c}$, their thermodynamic behavior is governed by critical exponents taking simple mean-field values. In contrast to the simplicity of the thermodynamic behavior, the behavior of physical models on \textit{finite} lattices is surprisingly subtle, and has been the subject of extensive numerical and theoretical studies; see e.g.~\cite{Brezin1982,Fisher1983,BinderNauenbergPrivmanYoung1985,Binder1985,RickwardtNielabaBinder1994,BercheKennaWalter2012,KennaBerche2014, WittmannYoung2014,FloresSolaBercheKennaWeigel2016,LundowMarkstrom2014,LundowMarkstrom2016,GrimmElciZhouGaroniDeng2017,Grimm2018,ZhouGrimmFangDengGaroni2018,caracciolo2001finite,jones2005finite,lv2019finitesize}.\par
In  recent years, the high-dimensional $(d \ge 5)$ finite-size scaling (FSS) behavior of the zero-field ferromagnetic Ising model on hypercubic lattices with either free (FBCs) or periodic boundary (PBCs) conditions has been of specific interest~\cite{BercheKennaWalter2012,KennaBerche2014, WittmannYoung2014,FloresSolaBercheKennaWeigel2016,LundowMarkstrom2014,LundowMarkstrom2016,GrimmElciZhouGaroniDeng2017,Grimm2018,ZhouGrimmFangDengGaroni2018}. With FBCs at the infinite-volume critical point $K_c$, it has been observed~\cite{LundowMarkstrom2014,GrimmElciZhouGaroniDeng2017} that the susceptibility $\chi$ and the two-point function $g(\mathbf{x})$ scale as $\chi \sim L^2$ and $g(\mathbf{x}) \sim \| \mathbf{x} \|^{2-d}$ where $L$ is the linear system size and $\mathbf{x}$ is a site on the lattice. While these scaling behaviors agree with the prediction from mean-field theory (i.e.~the Gaussian fixed point), it has, with PBCs, been observed~\cite{BercheKennaWalter2012,WittmannYoung2014,GrimmElciZhouGaroniDeng2017} that $\chi \sim L^{d/2}$ and $g(\mathbf{x}) \sim c_1\| \mathbf{x} \|^{2-d} + c_2L^{-d/2}$ where $c_1,c_2>0$. \par 

The surprising role of boundary conditions in high dimensions was recently addressed by taking various geometric approaches. In~\cite{GrimmElciZhouGaroniDeng2017}, it was suggested that the apparent breakdown of standard FSS with PBCs can be elucidated by studying the Ising model in its high-temperature representation. It was numerically observed that the critical PBC behavior is a manifestation of the proliferation of windings, absent for dimensions below $d_c$, and can be repaired by considering an appropriate definition of length, which accounts for the number of such windings. Another study~\cite{ZhouGrimmFangDengGaroni2018} investigated the boundary dependent FSS behavior of the Ising two-point function by introducing an appropriate random walk model. It was argued there that, above $d_{\rm c}=4$, this model displays the same universal FSS behavior as the Ising model.\par

Motivated by the fruitful insights from geometric approaches to high-dimensional FSS, we study here another well-known geometric representation of the Ising model, the $q$-state Fortuin-Kasteleyn (FK) random cluster model~\cite{Grimmett2006} with $q=2$. The random cluster model on a finite connected graph $G=(V,E)$ is defined by randomly choosing a spanning subgraph $(V,A)$ via the probability
$$
\pi(A) = \frac{1}{Z(p,q)}p^{|A|}(1-p)^{|E\backslash A|}q^{c(A)},
$$
where $p$ relates to the coupling constant $K$ via $p = 1 - e^{-2K}$, $c(A)$ is the number of connected components in $A$, and the partition function is given by $Z(p,q) = \sum_{A\subseteq E} \pi(A)$.\par

In this paper, we provide a detailed numerical study of the FSS behavior of a number of quantities in the FK Ising model on five-dimensional ($d=5$) hypercubic lattices with PBCs. At the estimated value of the infinite-volume critical point, we observe that the susceptibility $\chi:=L^{-d}\sum_i \langle \mathcal{C}_i^2 \rangle$, where $\mathcal{C}_i$ is the $i$-th largest cluster, scales as $L^{d/2}$. Note that this definition is equivalent to $\chi: = L^{-d}\text{Var}(M)$, where $M$ is the total magnetization in the spin representation of the Ising model. Moreover, we observe that the average size of the largest cluster $\langle \mathcal{C}_1 \rangle$ scales as $L^{3d/4}$. Using the expected scaling forms $\chi \sim L^{2\yhstar - d}$ and $\langle \mathcal{C}_1 \rangle \sim L^{\yhstar}$, both results suggest that the fractal dimension, which is equal to the magnetic exponent, takes the value $y_{\rm h}^{*} = 3d/4$, in agreement with the rigorously known value~\cite{LuczakLuczak2008} on the complete graph $K_n$, by setting $n:=L^d$, and in line with the numerical observation in~\cite{lundow2015complete}. However, in contrast to the scaling behaviors for $\chi$ and $\langle \mathcal{C}_1 \rangle$, we observe that the \textit{reduced} susceptibility $\chi^\prime:=\sum_{i \neq 1} \langle \mathcal{C}_i^2 \rangle / L^d$, i.e.~the contribution of all remaining clusters apart from the largest cluster, is bounded by $L^2$. More precisely, our data suggest that the scaling behavior can be well described by $\chi^\prime \sim L^2/\ln L$. Using the expected scaling form $\chi^\prime \sim L^{2\yh -d}$, this implies that $\yh = 1 + d/2$, in agreement with the prediction from the Gaussian fixed point. Moreover, we observe that, when removing the contribution of the largest cluster, the size dependence of the mean radius of gyration $R^\prime(s)$ scales as $R^\prime \sim s^{1/\yh}$, with $\yh \approx 7/2$ (for a precise definition see Sec.~\ref{Observables and simulation}). These observations are in agreement with the prediction from the Gaussian fixed point, suggesting $y_{\rm h} = 1+d/2$. \par
In the critical window around $K_c$, our simulations suggest that $\langle \mathcal{C}_1 \rangle \sim L^{3d/4} \tilde{C}_1(t L^{d/2})$, in agreement with the prediction from the complete graph and the value $y_{\rm t}^{*} = d/2$. Here $t = (K_{\rm c}-K)/K_{\rm c}$ is the reduced temperature and $\tilde{C}_1(\cdot)$ is the scaling function. For $\chi^\prime$, the FSS behavior depends on whether one approaches the critical point from the high-temperature ($t\rightarrow 0^+$) or low-temperature phase ($t\rightarrow 0^-$). In the case $t\rightarrow 0^+$, we observe $\chi^\prime \sim L^2\tilde{\chi}^\prime_{\rm h}(t L^{2})$, implying $\yt = 2$ and $\yh = 1 + d/2$, in agreement with the prediction from the Gaussian fixed point and a critical window of size $O(L^{-2})$. However, in the case $t\rightarrow 0^-$, our data suggest $\chi' \sim (\ln L)^{-1} L^{2} \tilde{\chi}^\prime_{\rm l}(-t L^{d/2})$; that is, we observe a logarithmic correction and a smaller critical window of size $O(L^{-d/2})$ compared to the high-temperature case. Here, $\tilde{\chi}^\prime_{\rm l}(\cdot)$ and $\tilde{\chi}^\prime_{\rm h}(\cdot)$ are scaling functions.\par

 We also numerically confirm the scaling picture from above by studying the FSS behavior of the cluster-size distribution and its reduced version; see Sec.~\ref{Results} for a detailed description. Our central result is therefore as follows: For a complete description of the PBC behavior of the five-dimensional FK Ising model, the scaling predictions from both the complete graph and the Gaussian fixed point need to be taken into account. More specifically the five-dimensional FSS behavior of observables which exclude the contribution of the largest cluster are mainly controlled by the Gaussian fixed point, where $y_{\rm t} =2$ and $y_{\rm h} =1+d/2$. However, observables which include the contribution of the largest cluster exhibit complete graph asymptotics, with $y_{\rm t}^{*}=d/2$ and $y_{\rm h}^{*}=3d/4$, consistent with the rigorously known values~\cite{LuczakLuczak2008} on the complete graph $K_n$.\par

The remainder of this article is organized as follows. In Sec.~\ref{Observables and simulation}, we define the investigated observables and state our simulation details. Our numerical results are presented in Sec.~\ref{Results}. Finally, we conclude with a discussion in Sec.~\ref{Discussion}.

  \section{Observables and simulation details}
  \label{Observables and simulation}
  
  We simulate the FK Ising model via the Swendsen-Wang (SW)~\cite{swendsen1987nonuniversal} and Wolff algorithm~\cite{wolff1989collective}. The update scheme of the SW algorithm can be described as follows. For a given bond configuration, identify all connected components and for each of them, randomly and uniformly assign either $+$ or $-$ spins to every vertex. This step maps configurations from bonds to spins. For a given spin configuration, proceed as follows: For each vertex, add a bond between this vertex and its adjacent vertices with probability $p = 1 - e^{-2K}$, where $K$ is the coupling constant if they have the same spin; otherwise, do not add a bond. \par 
 
  In the Wolff algorithm, the update scheme is as follows: Given a spin configuration, grow only one cluster from a uniformly and randomly chosen vertex, using the same connectivity rule as for SW. Then, generate a new spin configuration by flipping all spins on the cluster. \par
  
  In two and three dimensions, it was numerically observed~\cite{wolff1989comparison} that the Wolff algorithm has a smaller dynamic exponent than SW. Thus, in simulations, we used the Wolff algorithm to update and decrease the correlations of spin configurations and used the SW update only to create clusters for sampling. The number of Wolff updates between two consecutive SW steps is chosen to be approximately the volume divided by the averaged size of clusters, such that every spin has a decent chance to be updated during these steps. In both Wolff and SW updates, to speed up the process of adding bonds between adjacent vertices with the same spin, we adopt the procedure described in detail in~\cite{huang2018critical}. \par
  
 In simulations, each time the SW step maps configurations from spins to bonds, the following observables are sampled:
  \begin{enumerate}[label=(\roman*)]
      \item The first is the size ${\mathcal C}_1$ of the largest cluster. Moreover, we sample $\sum_{i\neq 1}{\mathcal C}^2_i$ where ${\mathcal C}_i$ is the size of the cluster $i$. The sum is over all clusters other than the largest cluster.
      \item For a cluster $C$, its radius of gyration ${\mathcal R}(C)$ is defined as
        \begin{equation}
        \label{Eq:Def_R(C)}
        {\mathcal R}(C) = \sqrt{\sum_{u\in C} \frac{({\bf x}_u - \bar{\bf x})^2}{|C|}},
        \end{equation}  
        where $\bar{\bf x} = \sum_{u\in C} {\bf x}_u/|C|$. Here ${\bf x}_u\in\ZZ^d$ is defined algorithmically as follows. First choose the vertex, say, $o$, in $C$ with the smallest vertex label, according to some fixed but arbitrary vertex labelling. Set ${\bf x}_o = {\bf 0}$. Start from the vertex $o$, and search through the cluster $C$ using breadth-first growth. Iteratively, ${\bf x}_v = {\bf x}_u + {\bf e}_i$ ($-{\bf e}_i$) if the vertex $v$ is traversed from $u$ along (against) the $i$-th direction, where ${\bf e}_i$ is the unit vector in the $i$-th direction. 
        
     \item $\mathcal{N}(s)$ is the number of clusters with size $[s,s+\Delta s)$.
    \item The average radius of gyration of clusters with size $[s,s+\Delta s)$, i.e.,
        \begin{equation}
          {\mathcal R}(s) := \frac{\sum\limits_{C:|C|\in[s,s+\Delta s)} {\mathcal R}(C)}{\mathcal{N}(s)},
        \end{equation}
       where we selected an appropriate bin size $\Delta s$ such that there were sufficient data in each bin.
  \end{enumerate}

We studied the ensemble averages of the following quantities: (\rmnum{1}) $C_1 = \langle \scrC_1 \rangle$, (\rmnum{2}) the susceptibility $\chi = L^{-d}\langle \sum_{i= 1}{\mathcal C}^2_i \rangle$ and its reduced version $\chi^\prime = L^{-d}\langle \sum_{i\neq 1}{\mathcal C}^2_i \rangle$, (\rmnum{3}) $R(s) = \langle {\mathcal R}(s) \rangle$ and its reduced version $R^\prime(s)$ where the largest cluster is removed, and (\rmnum{4}) the cluster-size distribution $n(s,L) = \frac{1}{L^{d} \Delta s}\left\langle \mathcal{N}(s) \right\rangle$ and its reduced version $n^\prime(s,L)$ where the largest cluster is removed.
    
At the estimated value of the infinite-volume critical point $K_{\rm c} = 0.1139150$~\cite{LundowMarkstrom2014}, we simulate systems with linear size $L= 4, 6, 8, 10, 12, 16, 20, 24, 28, 41, 51$. For $L =41$ and $L=51$, the sample sizes are $5 \times 10^5$ and $ 9 \times 10^5$, respectively. For each $L \leq 28$, no less than $10^6$ samples are generated. To further study the FSS behavior in the critical window, we also simulate systems with linear size $L= 16, 20, 24, 32, 40$. For each system size, we generate approximately $10^4$ samples. \par

We perform least-square fits on the FSS data. As a precaution against correction-to-scaling terms that we missed including in the fitting ansatz, we impose a lower cutoff $L \geq L_{\rm min}$ on the data points admitted in the fit, and systematically study the effect on the $\text{chi}^2$ value by increasing $L_{\rm min}$. In general, the preferred fit for any given ansatz corresponds to the smallest $L_{\rm min}$ for which the goodness of the fit is reasonable and for which subsequent increases in $L_{\rm min}$ do not cause the $\text{chi}^2$ value to drop by vastly more than one unit per degree of freedom. In practice, by `reasonable' we mean that $\text{chi}^2/{\rm DF} \lesssim 1$, where DF is the number of degree of freedom. The systematic error is estimated by comparing estimates from various sensible fitting ansatzes.
  
\section{Results}
\label{Results}

\subsection{FSS behavior at $K_c$}
\label{Section:fractal dimension}
\begin{figure}
         \centering
         \includegraphics[scale=0.7]{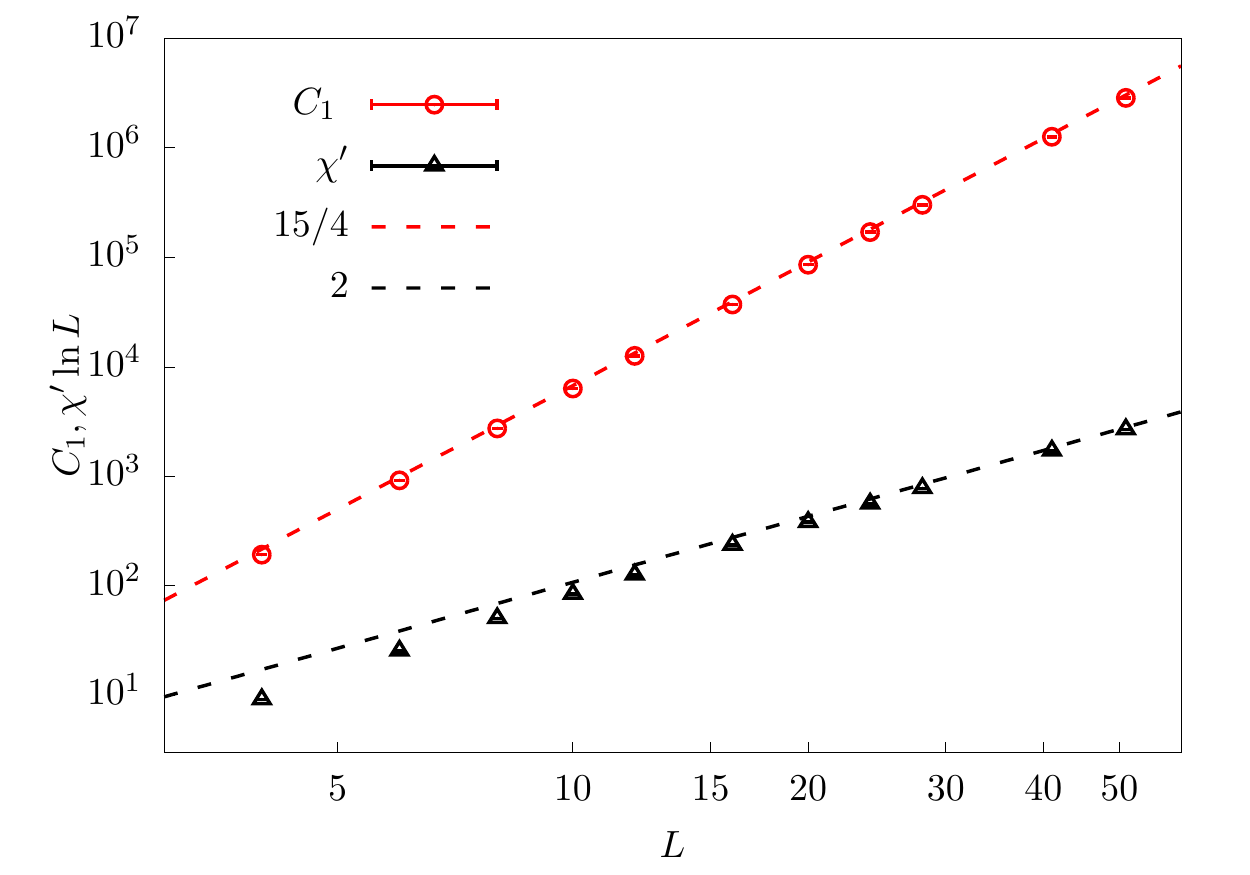}
         \caption{FSS behavior of the size of the largest cluster $C_1$ and the reduced susceptibility $\chi'$, which excludes the contribution of $C_1$, at $K_c$. The upper red line with slope $15/4$ corresponds to the prediction from the complete graph with $C_1 \sim L^{y_h^*} = L^{3d/4}$, while the lower black line with slope $2$ corresponds to the Gaussian fixed point prediction (accompanied by a multiplicative logarithmic correction) with $\chi'\ln L \sim L^{2\yh-d} = L^2$.}
         \label{Fig:C1 and the reduced chi}
       \end{figure}

In order to extract the magnetic exponent, We study the five-dimensional PBC behavior of $C_1$, $\chi^\prime$, $R^\prime(s)$, $n(s,L)$ and $n^\prime(s,L)$ at $K_c$. We first analyze the FSS behavior of $C_1$ in Fig.~\ref{Fig:C1 and the reduced chi}. We perform least-square fits using the ansatz
    \begin{equation}
      C_1 =  L^{y_{C_1}} (a_0 + a_1 L^{y_1} + a_2 L^{y_2} )\;,
    \end{equation}
where the exponent $y_{C_1}$ corresponds physically to the magnetic exponent and fractal dimension, and $a_1 L^{y_1}$ and $a_2 L^{y_2}$ are sub-leading corrections. Fixing $a_1 = a_2 = 0$ does not lead to stable fits, even for large $L_{\rm min}$. The same scenario holds when leaving $a_1, a_2, y_1, y_2$ as free parameters. However, fixing $a_2=0$ and leaving $a_1, y_1$ as free parameters, gives reasonable fits when $L_{\rm min} = 8$. We also vary the values of $y_1, y_2$, and study the effect on the estimate of $y_{C_1}$. Based on the fits, our final estimate is $y_{C_1} = 3.74(1)$. This exponent value is in agreement with $\yh^{*}  = 3d/4$ as suggested by the asymptotics on the complete graph, and in agreement with the numerical observation in~\cite{lundow2015complete}. We summarize our fitting results in Table~\ref{Tab:C1}.\par
       \par
     \begin{table}
           \centering
           \scalebox{0.9}{
           \begin{tabular}{lcccccccc}
           \hline
           $L_{\rm min}$ & $a_0$ & $a_1$ & $a_2$ & $\yh^*$ & $y_1$ & $y_2$ & $ \text{chi}^2/{\rm DF}$ \\
           \hline
%           6 & 1.149(5) & -1.5(5)   & 0   & 3.746(1) &-2.2(2) & 0 & 7.0/6 \\ 
           8 & 1.18(3)  & -0.4(2) & 0   & 3.739(6) &-1.2(5) & 0 & 3.1/5 \\ 
           10& 1.19(7)  & -0.3(3) & 0   & 3.74(1)  &-1(1)   & 0 & 3.1/4 \\ 
           \hline
           8 & 1.19(2)  & -0.2(2) &-0.2(6) &3.738(5)& -1  &-2 & 3.1/5 \\
           10& 1.19(4)  & -0.3(3) &-0.1(13)&3.738(7)& -1  &-2 & 3.1/4 \\
           \hline
            \end{tabular}}
           \caption{Fitting results for $C_1$. Some values do not have error bars since we fixed a specific value for these parameters in the ansatz.}
           \label{Tab:C1}
         \end{table}\par
  
      \begin{table}
 	\centering
 	\scalebox{0.9}{
 		\begin{tabular}{lcccccccc}
 			\hline
 			$L_{\rm min}$ & $a_0$ & $a_1$ & $a_2$ & $y_{\chi'}$ & $y_1$ & $y_2$ & $ \text{chi}^2/{\rm DF}$ \\
 			\hline
 			8 & 0.08(3)  & -1.02(3)  & -0.9(1)   & 2.06(4)  &-1/2 & -1 & 3.6/5 \\ 
 			10& 0.08(7)  & -1.02(6)  & -0.9(2)   & 2.06(8)  &-1/2 & -1 & 3.6/4 \\
 			\hline
 	\end{tabular}}
 	\caption{Fitting results for $\chi'$. Some values do not have error bars since we fixed a specific value for these parameters in the ansatz.}
 	\label{Tab:S2_1}
 \end{table}\par

      \begin{table}
	\centering
	\scalebox{0.9}{
		\begin{tabular}{lcccccccc}
			\hline
			$L_{\rm min}$ & $a_0$ & $a_1$ & $a_2$ & $y_{\chi^\prime}$ & $y_1$ & $y_2$ & $ \text{chi}^2/{\rm DF}$ \\
			\hline 
			8 & 1.3(1)   & -1.80(6)  & 0   & 1.96(1)  &-0.61(7)& 0 & 4.1/5 \\ 
			10& 1.3(2) & -1.78(3)  & 0   & 1.98(2)  &-0.7(1) & 0 & 3.6/4 \\ 
			12& 1.1(2) & -1.9(2)   & 0   & 1.99(3)  &-0.8(2) & 0 & 3.3/3 \\
			\hline
			6 & 1.089(8) & -2.71(6)  & 2.7(2) &1.999(2)& -1  &-2 & 3.7/6 \\
			8 & 1.08(2)& -2.7(2) & 2.5(4)   &2.000(3)& -1  &-2 & 3.5/5 \\
			10& 1.08(2)  & -2.7(3)	 & 2.5(10)  &2.000(6)& -1  &-2 & 3.5/4 \\
			\hline
	\end{tabular}}
	\caption{Fitting results for $\chi' \ln L$. Some values do not have error bars since we fixed a specific value for these parameters in the ansatz.}
	\label{Tab:S2_2}
\end{table}

For the reduced susceptibility $\chi'$, we perform least-square fits to the ansatz $\chi' = L^{y_{\chi'}}(a_0 + a_1L^{y_1} + a_2L^{y_2})$. Including only one correction term, i.e.~$a_1L^{y_1}$, and leaving $y_1$ as a free parameter leads to unstable fits. Including both correction terms and setting $y_2=2y_1$ and $y_1=1/2$ gives $y_{\chi'} = 2.06(8)$ (see Table~\ref{Tab:S2_1}). However, these fits are not convincing since $a_1$ is much larger than $a_0$, implying that the correction term is bigger than the leading term. Moreover, the extracted value for $a_0$ is not convincing, since its error bar is comparable to its estimated value. We note that the scaling of the second largest cluster on the complete graph has a logarithmic factor~\cite{LuczakLuczak2008}. We therefore conjecture that a logarithmic correction for $\chi^\prime$ also appears on five-dimensional hypercubic lattices which we numerically confirm now. Our fits onto the alternative ansatz $\chi' \ln L = L^{y_{\chi'}} (a_0 + a_1L^{y_1} + a_2L^{y_2})$ lead to stable estimates of $y_{\chi^\prime}=2$ (see Table~\ref{Tab:S2_2} for details). Using the expected scaling form $\chi' \sim L^{2y_{\rm h} -d }$, implies $\yh=1+d/2$, in agreement with the prediction from the Gaussian fixed point. \par

\begin{figure}
         \centering
         \includegraphics[scale=0.7]{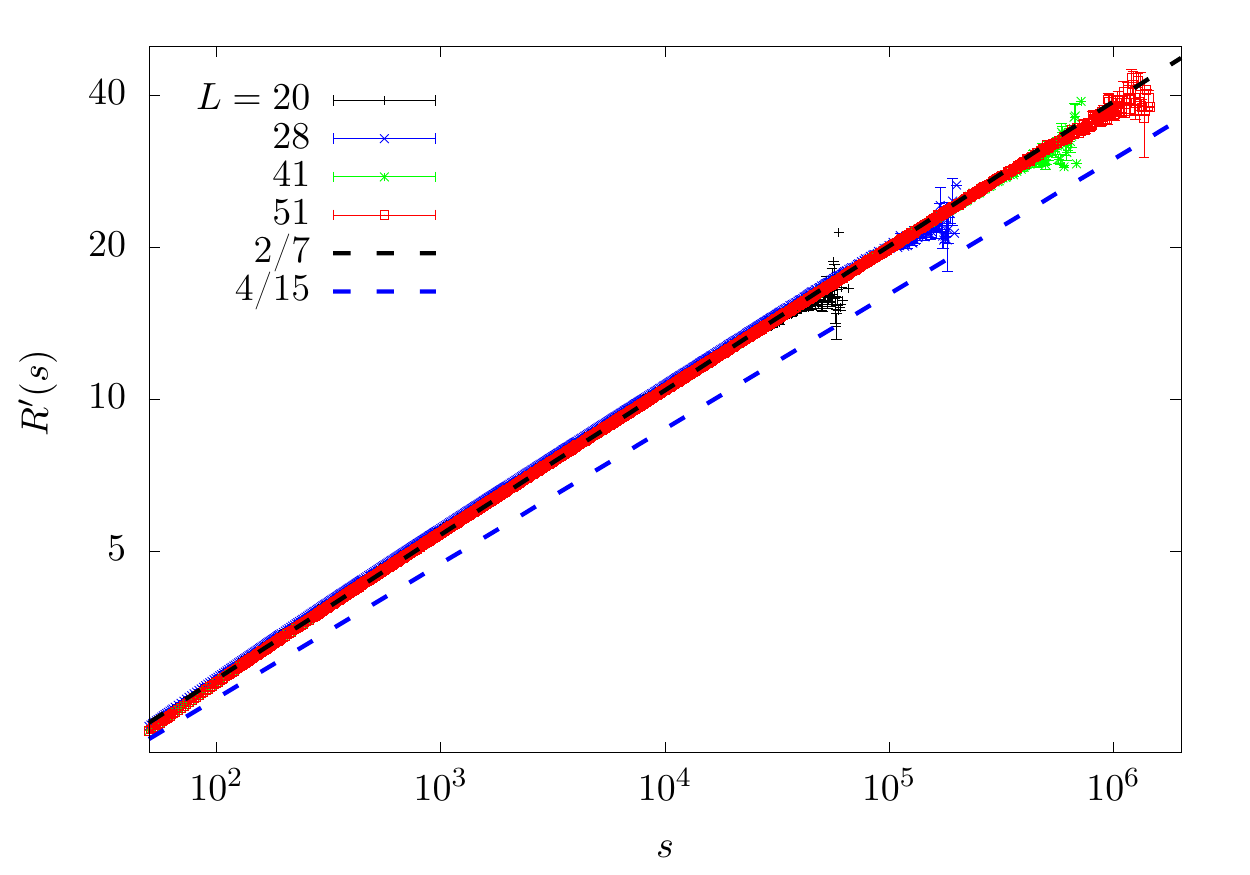}
         \caption{FSS behavior of the reduced radius of gyration, $R^\prime(s)$, where the contribution of the largest cluster is removed, at $K_{\rm c}$. Our data suggest the scaling $R^\prime(s) \sim s^{1/y_{\rm h}} = s^{2/7}$, in agreement with the prediction of the Gaussian fixed point. The upper black line with slope $2/7$ corresponds to the scaling asymptotics from the Gaussian fixed point. The lower blue line with slope $4/15$ corresponds to the predicted scaling from the complete graph.}
         \label{Fig:radius of gyration}
       \end{figure}

We now turn to $R^\prime(s)$. The size-dependence of the radius of gyration is characterised by the fractal dimension, or magnetic exponent, as $R^\prime(s) \sim s^{1/\yh}$. Figure~\ref{Fig:radius of gyration} shows $R^\prime$ against the cluster size $s$. Our data clearly suggest that $\yh$ is consistent with $1+d/2$. For completeness, we also measure $R$ for the largest FK cluster. By following a fitting approach similar to that for $C_1$, we obtain $R \sim L^{0.99(1)}$, in agreement with $C_1 \sim L^{\yh^*} \sim R^{\yh^*}$.\par
We also studied the cluster-size distribution $n(s,L)$. The standard FSS formula is given by
\begin{equation}
    n(s,L) \sim s^{-\tau}\tilde{n}\left(s/L^{\yh^*}\right),
\end{equation}
where $\tilde{n}(\cdot)$ is a universal scaling function and $\tau$ is the Fisher exponent. The scaling function $\tilde{n}(x)$ is roughly a constant when $x \ll 1$ and then decays quickly to zero when $x\gg 1$. In other words, the standard FSS behavior of $n(s,L)$ first exhibits a power-law behavior with exponent $\tau$, and then starts to decay around $s\sim L^{\yh^*}$. Moreover, as numerically observed for percolation models in various dimensions~\cite{hoshen1979monte,huang2018critical} and for the two- and three-dimensional FK Ising model ~\cite{ruge1992scaling,ruge1993study,de1990monte,hou2019geometric}, the equation
\begin{equation}
\label{eq:scaling_relation}
\tau = 1 + d/\yh^*
\end{equation}
is expected to hold. Figure~\ref{Fig:n(s,L)} shows $n(s,L)$ against $s$. Initially, $n(s,L)$ displays a power-law behavior, then enters a plateau and, finally, decays significantly. The slope of the dashed line at the bottom is consistent with the value $17/7$. Applying $\tau=17/7$ to Eq.~\eqref{eq:scaling_relation} leads to an exponent $7/2$ which disagrees with $\yh^*$ but is in agreement with $\yh$. However, the upper dashed line in Fig.~\ref{Fig:n(s,L)}, which is defined by connecting the end points of the plateau for each system size $L$, displays a power-law behavior with an exponent value of $7/3$ rather than $17/7$. Applying $\tau = 7/3$ to Eq.~\eqref{eq:scaling_relation}, leads to an exponent of $15/4$, in agreement with $\yh^{*}$.\par

       \begin{figure}
         \centering
         \includegraphics[scale=0.7]{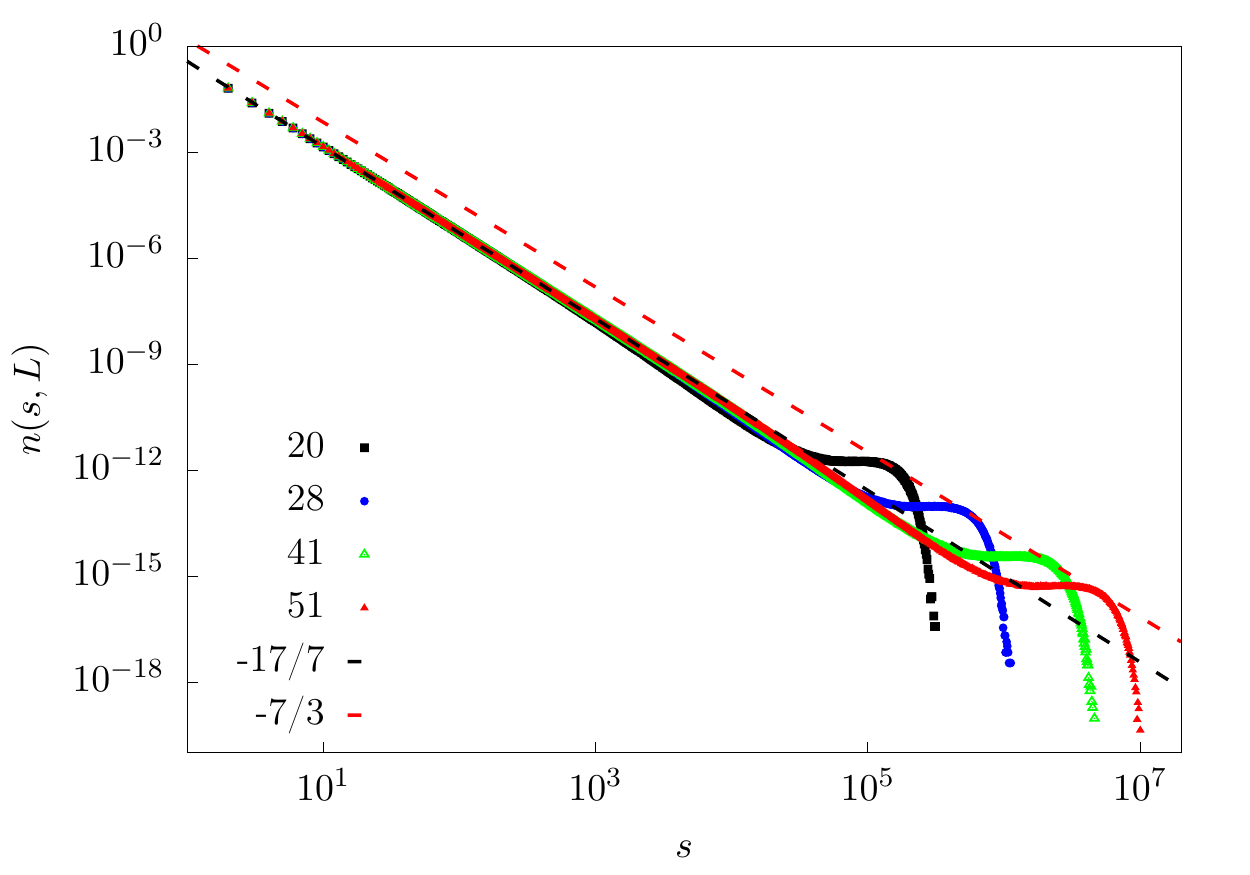}
         \caption{FSS behavior of the cluster-size distribution $n(s,L)$ at $K_{\rm c}$. The upper dashed line with slope $-7/3$ corresponds to the scaling prediction from the complete graph, while the lower dashed line with slope $-17/7$ corresponds to the prediction from the Gaussian fixed point.}
         \label{Fig:n(s,L)}
       \end{figure}
       
       \begin{figure}
         \centering
         \includegraphics[scale=0.7]{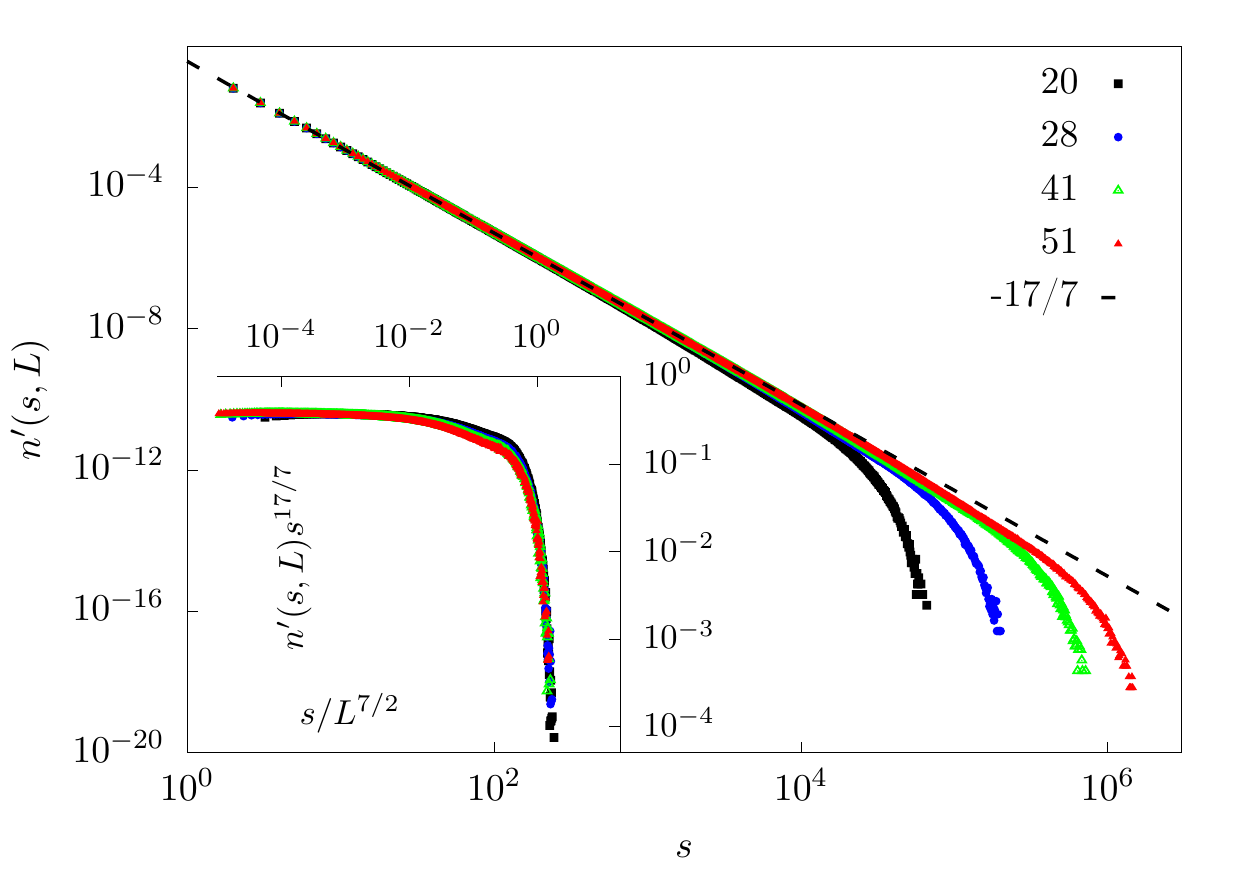}
         \caption{FSS behavior of the reduced cluster-size distribution $n'(s,L)$, where the contribution of the largest cluster is removed, at $K_{\rm c}$. The inset shows an appropriate scaled version of the data.}
         \label{Fig:n'(s,L)}
       \end{figure}
For the reduced cluster-size distribution $n'(s,L)$, we expect $n'(s,L) \sim s^{-\tau}\tilde{n}'\left(s/L^{\yh}\right)$ with $\tau = 1 + d/\yh$. Fig.~\ref{Fig:n'(s,L)} shows $n'(s,L)$ against $s$. $n'(s,L)$ clearly shows a power-law behavior with exponent value $\tau = 17/7$. Moreover, the inset of Fig.~\ref{Fig:n'(s,L)} shows an appropriately scaled version. The data collapse provides strong evidence of the correctness of the scaling formula for $n'(s,L)$. By comparing Figs.~\ref{Fig:n(s,L)} and~\ref{Fig:n'(s,L)}, plateaus appear only when taking the largest cluster into account.

\subsection{FSS behavior in the critical window}
     \begin{figure}
         \centering
         \includegraphics[scale=0.7]{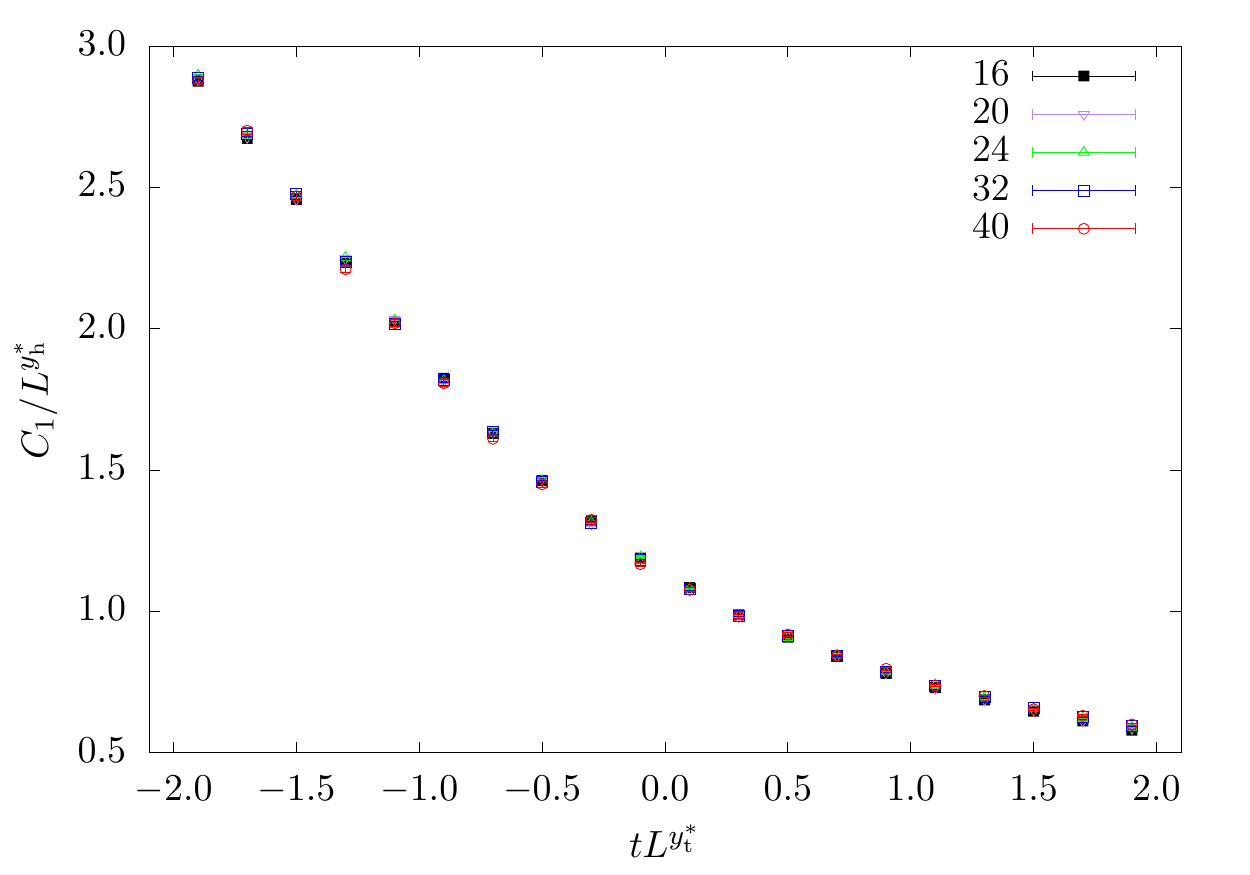}
         \caption{Appropriately scaled version of the FSS behavior of the largest cluster $C_1$ in the critical window close to $K_{\rm c}$. The data provide strong evidence in favor of the complete graph prediction.}
         \label{Fig:scalingwindow_C1}
       \end{figure} 
In order to extract the thermal exponent, we study the FSS behavior in the critical window close to $K_c$. We first consider the scaling behavior of $C_1$. On the complete graph, it is known rigorously~\cite{LuczakLuczak2008} that the size of the critical window is of the order $n^{-1/2}$, where $n$ is the number of vertices. Based on the observed FSS behavior of $C_1$ at $K_c$, where the largest cluster was observed to display complete-graph asymptotics, we conjecture the scaling formula $C_1(t,L) \sim L^{\yh^*}\tilde{C_1}(t L^{\yt^*})$ in a critical window of order $L^{-d/2}$, where $t = (K_{\rm c} - K)/K_{\rm c}$ and $\yt^* = d/2$. To test this conjecture, we chose, for each system size $L$ simulated, a coupling constant $K$ such that $tL^{y^{*}_{\rm t}} \in [-2,2]$. The excellent data collapse in Fig.~\ref{Fig:scalingwindow_C1} provides strong evidence for the correctness of our claim.\par

         \begin{figure*}[ht!]
         \centering
         \includegraphics[scale=0.71]{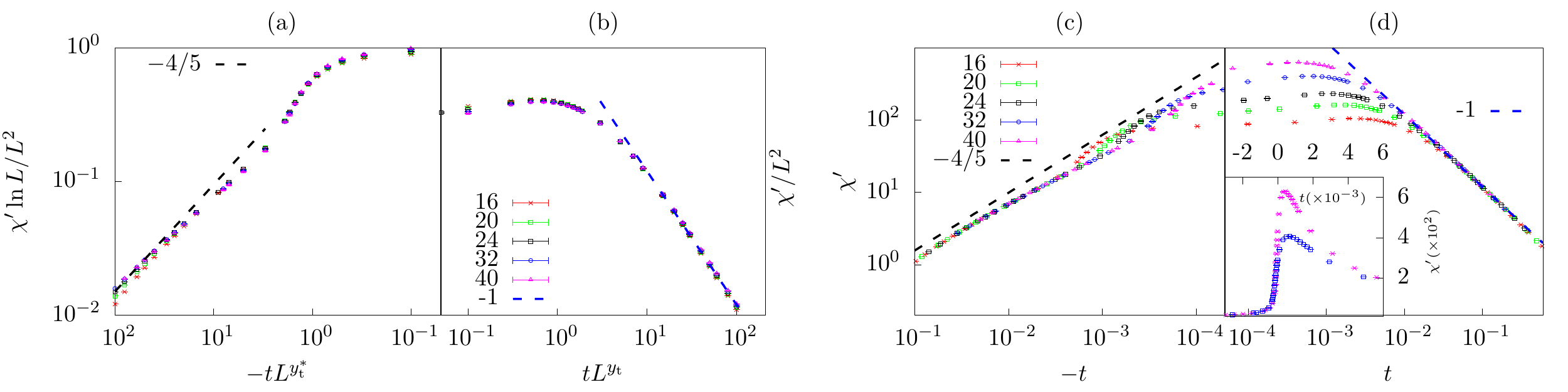}
         \caption{Left: FSS behavior of the reduced susceptibility $\chi^\prime$ in the ($a$) low- and ($b$) high-temperature critical windows. (Right) The behavior of $\chi^\prime$ in the thermodynamic limit when $(c)$ $t\to 0^-$ and $(d)$ $t\to 0^+$. The inset shows the data of $\chi^\prime$ close to the critical point, in which the divergence at $K_{\rm c}$ and the asymmetric behavior of $\chi^\prime$ close to $K_{\rm c}$ are apparent.}
         \label{Fig:scaling_chi'}
       \end{figure*}

Finally, we consider the FSS behavior of $\chi'$. In the case $t \to 0^+$, we plot $\chi^\prime(t,L)/L^2$ versus $tL^{2}$ in Fig.~\ref{Fig:scaling_chi'}$(b)$. The excellent data collapse suggests that $\chi^\prime(t,L) \sim L^{2\yh-d}\tilde{\chi}^\prime_{\rm h}(t L^{\yt})$ where $\yh = 1 +d/2$ and $\yt = 2$, in agreement with the prediction from the Gaussian fixed point. In order to extract the thermodynamic behavior $\chi^{\prime} \sim t^{-\gammaH}$, we study $\tildechiprimeH(x)$ with $x:=t L^{\yt}$ and
expect the scaling $\tildechiprimeH(x) \sim x^{-\gammaH}$ with $\gammaH = (2\yh -d)/\yt = 1$.
The scaling of $\tildechiprimeH(x)$ is demonstrated in Fig.~\ref{Fig:scaling_chi'}$(b)$, and the thermodynamic behavior $\chi^{\prime}(t\rightarrow 0^+,\infty)$ is directly confirmed in Fig.~\ref{Fig:scaling_chi'}$(d)$. In the case of $t \to 0^-$, Fig.~\ref{Fig:scaling_chi'}$(a)$ shows $\chi^\prime \ln L /L^2$ against $-tL^{5/2}$. The data collapse suggests $\chi^\prime(t, L)\sim (\ln L)^{-1}L^{2\yh - d} \tilde{\chi}^\prime_{\rm l}(-t L^{\ytstar})$, in agreement with $\yt^* = d/2$ as expected from the complete graph. As in the critical case, the multiplicative factor $\ln L$ also appears for $\chi^\prime(t,L)$ when $t<0$. In order to extract the thermodynamic behavior $\chi^{\prime} \sim (-t)^{-\gammaL}$ as $t\rightarrow 0^-$, we study $\tildechiprimeL(x)$ with $x:=-t L^{\ytstar}$ which is expected to scale as $\tildechiprimeL(x) \sim x^{-\gammaL}$ with $\gammaL = (2\yh -d)/\ytstar = 4/d$. The scaling of $\tildechiprimeL(x)$ is shown in Fig.~\ref{Fig:scaling_chi'}$(a)$, and the thermodynamic behavior $\chi^{\prime}(t\rightarrow 0^-,\infty)$ is confirmed in Fig.~\ref{Fig:scaling_chi'}$(c)$. Thus, in making a comparison with the $t\rightarrow 0^+$ case, we observe a smaller scaling window of size $O(L^{-d/2})$ as $t\rightarrow 0^-$. We assume the underlying reason is that when $t<0$ and outside of the critical window of size $O(L^{-d/2})$, the largest cluster becomes significantly large, such that $\chi^\prime$, which is from the contribution of all other clusters, is largely suppressed and becomes off critical. The asymmetric behavior of $\chi^\prime$ close to $K_{\rm c}$, i.e.~$\gammaH \neq \gammaL$, clearly appears in the inset of Fig.~\ref{Fig:scaling_chi'}(d).

  \section{Discussion}
  \label{Discussion}
  In this paper, we numerically studied the FSS behavior of a number of quantities of the Fortuin-Kasteleyn Ising model on five-dimensional hypercubic lattices with periodic boundary conditions. Our main results can be summarized as follows: Physical quantities which include the contribution of the largest FK cluster are seen to display complete graph asymptotics, in broad agreement with the numerical observations in~\cite{lundow2015complete}. However, the leading asymptotic behavior of observables, where the contribution of the largest cluster is removed, is no longer described by the complete graph but by the Gaussian fixed point prediction. For $d=5$, we therefore suggest that two sets of exponents, i.e.~$(y_t, y_h)$ and $(y_t^*,y_h^*)$, are needed to provide a consistent scaling picture for the PBC behavior of the FK Ising model. 
  
\begin{figure}
         \centering
          \includegraphics[scale=0.7]{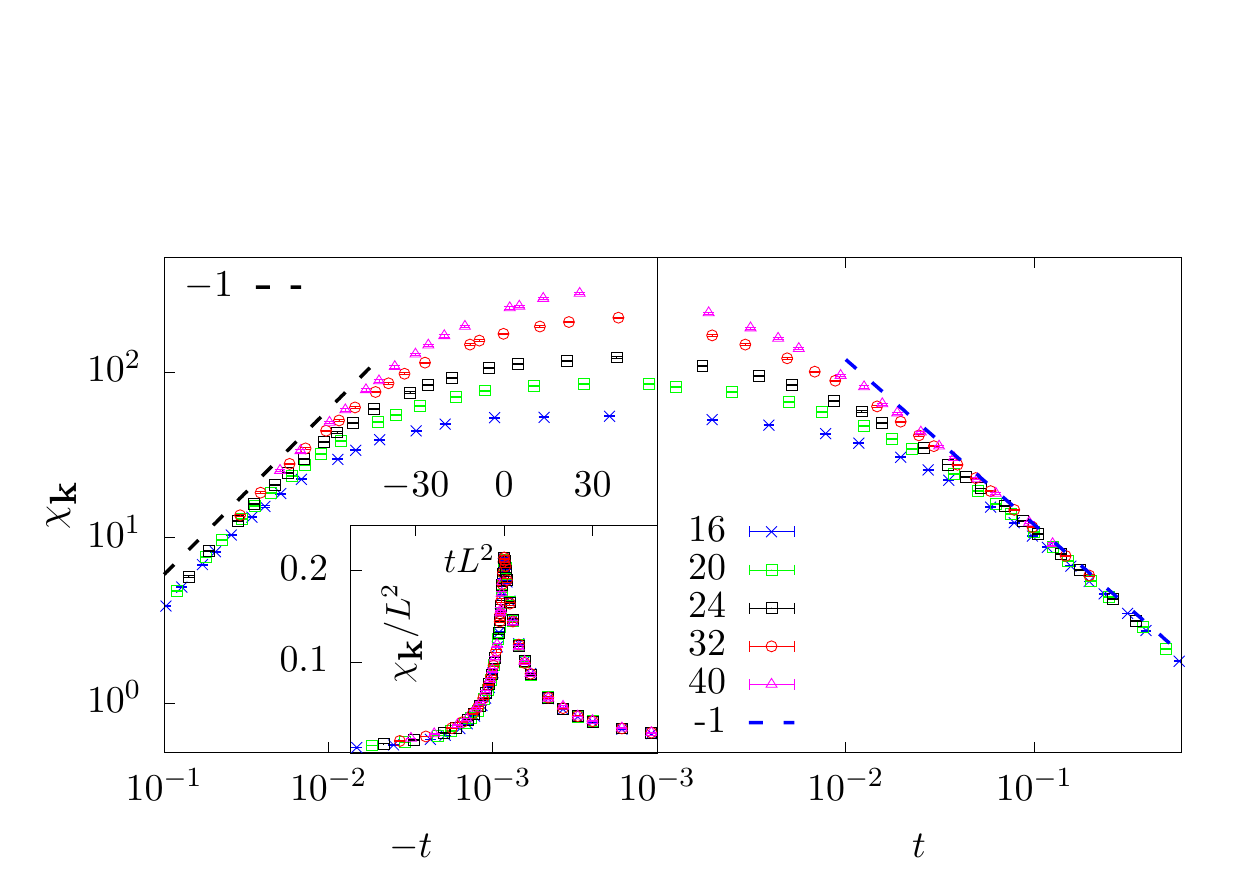}
         \caption{Plot of $\chi_{\bfk\neq 0}$ with $\bfk = (1,0,0,0,0)$ in low- (left) and high-temperature (right) regions. The inset plots a rescaled version, suggesting the finite-size scaling $\chi_{\bfk\neq 0}(t,L) \sim L^2 \tilde{\chi}_{\bfk \neq 0}(tL^2)$.}
         \label{Fig:chi_k}
\end{figure}

We note that, in~\cite{WittmannYoung2014}, the same two sets of critical exponents are observed, by studying the Fourier-transformed susceptibility $\chi_{\bfk}$ of the spin Ising model in five dimensions, with zero ($\bfk = 0$) and non-zero ($\bfk \neq 0$) mode;, see~\cite{WittmannYoung2014} for precise definitions. In particular, for fixed $\bfk \neq 0$, it was observed in~\cite{WittmannYoung2014}  that the finite-size scaling $\chi_{\bfk \neq 0}(t,L) \sim L^2\tilde{\chi}_{\bfk \neq 0}\left( t L^2\right)$, with $\tilde{\chi}_{\bfk \neq 0}(\cdot)$  being the scaling function, and in the thermodynamic limit, $\chi_{\bfk \neq 0}(t,\infty) \sim |t|^{-1}$ as $t\rightarrow 0$, as shown in Fig.~\ref{Fig:chi_k}. Thus, in the high-temperature region, $\chi_{\bfk \neq 0}$ and the reduced susceptibility $\chi^\prime$ exhibit the same finite-size scaling and infinite-volume behavior. However, in the low-temperature region, our data show the finite-size scaling of $\chi^\prime$ is consistent with $\chi^\prime \sim (\ln L)^{-1}L^2 \tilde{\chi}^\prime_{\rm l}\left(-t L^{d/2}\right)$ and $\chi^\prime \sim (-t)^{-4/5}$. We mention that the occurrence of the multiplicative logarithmic correction $(\ln L)^{-1}$ is merely a conjecture and very consistent with our numerical data.

These results motivate the question of whether similar scaling behaviors can be observed for the $O(n)$ vector model, where the case $n=1$ corresponds to the Ising model. For this model above $d_c$, it is well known, by using renormalization group arguments~\cite{fisher1983scaling}, that the PBC behaviour of the free energy density $f(t,h,L)$ is affected by dangerous irrelevant variables. In particular, it was proposed~\cite{BinderNauenbergPrivmanYoung1985,fisher1983scaling} that
  \begin{equation}
  f(t,h,L) \sim L^{-d} \tilde{f}_s (t L^{\tilde{y}_t}, hL^{\tilde{y}_h})
  \label{eq:free_energy}
  \end{equation} 
  where $\tilde{f}_s$ is the singular part of the scaling function, $h$ is the magnetic field, and $\tilde{y}_t$ and $\tilde{y}_h$ are modified thermal and magnetic exponents. It was argued, by taking the dangerous irrelevant scaling field $u$ associated with the $\phi^4$ term in the field-theoretical description into account, that these modified exponents satisfy $\tilde{y}_t =y_t-y_u/2$ and $\tilde{y}_h=y_h-y_u/4$ with $y_u=4-d$. By observation, these values coincide with the values of $y_t^*$ and $y_h^*$ from the complete graph prediction. Apparently, Eq.~\eqref{eq:free_energy} shows that the FSS behavior of $f(t,h,L)$ is solely controlled by the modified exponents $\tilde{y}_t$ and $\tilde{y}_h$. In other words, the scaling formula suggests that there is no need for \textit{additional} terms in Eq.~\eqref{eq:free_energy} which explicitly include the exponents $y_t$ and $y_h$. However, based on our observations for the five-dimensional FK Ising model, we conjecture that a complete description of high-dimensional PBC behavior requires more than just one magnetic and thermal exponent. In future, we will investigate the scaling formula for $f$, by numerically studying the high-dimensional PBC behavior of some special cases of the $O(n)$ vector model, and provide evidence in favor of our conjecture.
  
\section{Acknowledgments}
Y.D. acknowledges the support from the National Key R\&D Program of China under Grant No. 2016YFA0301604 and from the National Natural Science Foundation of China under Grant No. 11625522. J.G. thanks J. Chan for fruitful discussions. Z.Z. and S.F. thank the Research Support Scheme from ACEMS for providing financial support at Monash University, where this work was written. We also thank the Supercomputing Center of University of Science and Technology of China for the computer time.

\bibliographystyle{apsrev4-1}

\end{document}